\documentclass[twocolumn,amsmath,amssymb,pra,a4paper]{revtex4}
\usepackage{graphicx}
\usepackage{float}
\usepackage{epsfig}
\usepackage{subfigure}
\usepackage[utf8]{inputenc}
\usepackage[T1]{fontenc}
\usepackage{color}
\newcommand{\re}{\mathop\mathrm{Re}}
\newcommand{\im}{\mathop\mathrm{Im}}
\newcommand{\sech}{\mathop\mathrm{sech}}

\date{\today}
\begin{document}
\title{Frequency comb injection locking of mode locked lasers}
\author{Omri Gat}
\affiliation{Racah Institute  of  Physics, Hebrew University
of Jerusalem, Jerusalem 91904, Israel}
\author{David Kielpinski}
\affiliation{Centre for Quantum Dynamics, Griffith University, Nathan QLD 4111, Australia}
\begin{abstract}
The two-frequency problem of synchronization of the pulse train of a passively mode locked soliton laser to an externally injected pulse train is solved in the weak injection regime. The source and target frequency combs are distinguished by the spacing and offset frequency mismatches. Locking diagrams map the domain in the mismatch parameter space where stable locking of the combs is possible. We analyze the dependence of the locking behavior on the relative frequency and chirp of the source and target pulses, and the conditions where the relative offset frequency has to be actively stabilized. Locked steady states are characterized by a fixed source-target time and phase shifts that map the locking domain.
\end{abstract}
%\pacs{42.50}
\maketitle
\section{Introduction}
Synchronization, or injection locking of a self-sustained oscillator to an external periodic signal is a paradigm of nonlinear dynamics \cite{pik}. The frequency of the target oscillator adjusts itself to the external frequency for strong enough injection or small enough frequency mismatch, with a sharp threshold separating configurations with stable locked steady states from unlocked configuration. Injection locking in cw lasers has been extensively studied experimentally and theoretically \cite{rep}. Injection locking is useful to stabilize the target laser versus phase diffusion and frequency drift, and can be achieved for very weak seeding provided the frequency mismatch is small enough. Strong injection leads to rich behavior, including bifurcations, multistability, excitability and chaos \cite{rep}. 

If the laser is mode locked, and the injected signal is pulsed, the target laser pulse train can injection-lock to the source pulse train. Unlike cw synchronization that implies entrainment of a single optical frequency, pulse train synchronization entails entrainment of \emph{two} frequencies, namely the pulse repetition rate and the pulse phase shift per round trip. Since these two frequency determine the spacing and offset of the pulse train frequency comb, pulse train synchronization is equivalent to injection locking of the target laser frequency comb to the comb of the source. In this way a very high-quality standard frequency comb source, that can be quite weak, could impart its quality to the target laser, locking it to the standard and reducing imperfections, such as pulse noise and frequency drift. A natural application is when the source is carrier-envelope phase locked, where injection locking locks the carrier-envelope phase of the target.
When the repetition rates of the source and target are not matched but rationally related, synchronization results with a target repetition rate that is the least common multiple of the source and free running target, facilitating generation of high-repetition rate frequency comb \cite{kg12}.

Compared to cw injection locking, there are few studies of injection locking of mode locked lasers. The basic theory of the system was developed in \cite{moh1} using soliton perturbation theory in the framework of the Haus mater equation model. Reference \cite{moh1} derived estimates of the locking regime parameters, and studied numerically the pulse dynamics for some typical locked and non-locked parameters. The phenomenon was demonstrated experimentally in actively mode locked lasers in \cite{mikael} and in passively mode locked fiber lasers in \cite{mo,dorring}. The effects of noise in injection locked mode locked lasers was studied in \cite{moh2}. The combined effect of noise and injection in the mode locking transition was studied theoretically in \cite{crit-th} and experimentally in \cite{crit-exp}, showing critical phenomena with classical exponents. Rational injection locking was demonstrated experimentally in passively mode locked lasers in \cite{moe,kg12} and in harmonic mode locked lasers in \cite{wang,kurita}.

Here we focus on the injection locking \emph{threshold} phenomenon. If the envelope of a pulse of the free-running target laser is $\psi_\text{fr}(t)$, its pulse train waveform is
\begin{equation}\label{eq:fr}
\psi_\text{free-running}=\sum_n\psi_\text{fr}(t-n\tau_r)\,e^{in\nu_0\tau_r}
\end{equation}
where $\tau_r$ is the repetition rate, and $\nu_0\tau_r$ is the pulse to pulse overall phase shift. Assuming that the source wave form is
\begin{equation}\label{eq:so}
\psi_\text{source}=\sum_n\psi_\text{so}(t-n(1+u)\tau_r)\,e^{in(\nu_0+\nu)\tau_r}
\end{equation}
where $\psi_\text{so}(t)$ is the source pulse wave form, the time slip rate $u$ and phase slip rate $\nu$ between the source and the target pulse trains measure the mismatch between the source and target frequency combs. Since the pulse to pulse phase shift is defined modulo $2\pi$, $\nu$ is defined modulo $\frac{2\pi}{\tau_r}$, while the frequency $u$ can take all real values.

In standard single-frequency synchronization there is a threshold frequency mismatch below which the target is locked. At threshold the locked state loses stability in a saddle-node bifurcation. Here there are two mismatch parameters, $u$ and $\nu$, so for a given $\psi_\text{so}(t)$ there is a domain in the $\nu,u$ plane of where the target frequency comb is stably injection locked to the source comb. In the locking domain the target waveform is then
\begin{equation}\label{eq:il}
\psi_\text{injection-locked}=\sum_n\psi_\text{il}(t-n(1+u)\tau_r)\,e^{in(\nu_0+\nu)\tau_r}
\end{equation}
where the $\psi_\text{il}$ is the target pulse waveform in the presence of injection.

The shape and area of the locking domain has to be calculated on a case-by-case basis. Here we study soliton mode locking where the free target pulse waveform is non-chirped soliton-like $\psi_\text{fr}(t)=a\sech(at)$, and  the source pulse waveform is 
\begin{equation}\label{eq:as}
\psi_\text{so}(t)=\kappa a\sech(at)e^{i\omega t+\beta t^2}
\end{equation}
so that the injected pulse shape is an attenuated version of the free-running pulse shape; still, it follows from the analysis below that the locking properties depend weakly on the precise source pulse shape as long as it approximately matches the target pulse.

The parameters of the source pulse are the dimensionless injection strength $\kappa$, frequency $\omega$ and chirp $\beta$. The parameter $\omega$ is usually omitted from the definition of the pulse shape because it can be absorbed into the carrier frequency. Here it has to be spelled out explicitly because the difference between the peak gain frequencies of the source and target pulses is a crucial locking parameter. %There is a subtle difference between $\omega$ and the difference $\Delta\omega_c$ between the carrier frequencies of the source and target: Supposing that the target waveform (equation \ref{eq:fr}) and the source waveform (equation \ref{eq:so}) are specified with respect to carrier frequency $\omega_c$ and $\omega_c+\Delta\omega_c$ (respectively), the carrier frequency of the source can be shifted to $\omega_c$ by changing $\psi_\text{so}\to\psi_\text{so}e^{i\Delta\omega_ct}$ and $\nu\to\nu+\Delta\omega_c\tau_r$. We choose to parametrize the frequency gap using $\omega$ rather than $\Delta\omega_c$ to streamline the theory.

In most of our analysis we assume that the injection is weak, $\kappa\ll1$. As shown below, under this assumption the locked steady state and its stability depend only on the ratios $\frac{\nu}{\kappa}$ and $\frac{u}{\kappa}$ and the main effect of the injection on the target waveform is a shift $\Delta$ in its overall timing and a shift $\delta$ in its overall phase relative to the source pulse, so that $\psi_\text{il}\approx a\sech(a(t-\Delta))e^{i\delta}$.

The main results of this work are the locking diagrams shown in figure \ref{fig:ld} for several values of pairs $\omega,\beta$. The boundaries of the locking domains are shown in black. In addition to the threshold, the locking diagrams also show the $\Delta$ and $\delta$ level curves. Interestingly, although in general the locking domains are smaller for larger $|\omega|$ and $|\beta|$ due to decreasing source-target pulse overlap, the shrinking is not uniform, and locking with some mismatch parameters is possible only for nonzero relative frequency or chirp. When $\omega\ne0$ there is a correlation between the timing and phase frequency mismatches, where for some ${\nu}$ values there is a minimal $u$ needed for synchronization and vice versa.

\section{The injection locking master equation}
Our analysis is based on the Haus theory of anomalous dispersion passively mode-locked lasers with a fast saturable absorber \cite{haus}, also called mode locked soliton lasers \cite{kutz}. The Haus master equation is a multiple scale model of the electric field envelope $\psi(\tau,t)$, where $t$ is the fast sub-roundtrip time scale, and $\tau$ is the slow time scale describing the evolution of the wave form for an integer multiple of the roundtrip time $\tau_r$.

Using the gain band center as the reference frequency the equation of motion for the free target laser is
\begin{equation}\label{eq:master}
\partial_\tau\psi=i(\partial_t^2+2|\psi|^2)\psi+g[\psi]\left(1+\gamma\partial_t^2\right) \psi+ \sigma(|\psi|^2)\psi
\end{equation}
where $g$ is the overall gain, assumed to depend on the pulse energy $\int dt|\psi|^2$, $\gamma$ is the coefficient of parabolic spectral gain profile, and  $\sigma$ is the transmissivity of the saturable absorber, including linear loss equal to $-\sigma(0)$. The first two terms on the right-hand-side of equation (\ref{eq:master}) model the anomalous dispersion and Kerr nonlinearity, with coefficients fixed by appropriate choice of units. The conversion of the results to physical units and their implication for practical systems are discussed below.

Since the Haus master equation describes the evolution of the pulse waveform between consecutive roundtrip periods, the optical injection can be modeled by an inhomogeneous seed term $\psi_\text{ih}(t,\tau)$, where $\psi_\text{ih}(t,n\tau_r)$ is the envelope of the seed pulse waveform (in appropriate units) injected in the $n$th roundtrip in the frame defined by equation (\ref{eq:master}). For the source waveform (\ref{eq:so}) the injection locking master equation is therefore
\begin{align}\nonumber
\partial_\tau\psi=i(\partial_t^2+2|\psi|^2)\psi
&+g(1+\gamma\partial_t^2) \psi\\&+ \sigma(|\psi|^2)\psi+\psi_\text{so}(t-u\tau)e^{i(\nu_0+\nu)\tau}
\label{eq:master-inj}\end{align}
The injection locking master equation that was first derived in \cite{moh1} is equivalent to equation (\ref{eq:master-inj}). The master equation was written in  \cite{moh1} in the frame of reference where the injected signal is exactly periodic, and therefore contains an additional term in the master equation that arises from a frequency shift of the source waveform, that is accounted here by the parameter $\omega$.

The target waveform approaches
an injection locked pulse train (\ref{eq:il}) if there is a stable solution of Eq.\ (\ref{eq:master-inj})  of the form
$\psi_\text{il}(t-u\tau)e^{i(\nu_0+\nu)\tau}$. 
 An unlocked solution typically consists of two pulses, a strong pulse of free target waveform weakly perturbed by interaction with the seed, and a weak response waveform that is approximately a linearly filtered version of the seed pulse. The unlocked state is not stationary in any frame.

%The injected pulse waveform is largely arbitrary other than being localized. Nevertheless, the size and shape of the injection locking region in the $w\nu$ plain depends strongly on the precise form of $\phi$. In particularly better synchronization properties are expected when the seed and target pulses are well-matched in width, frequency and chirp.

\section{Soliton perturbation theory}
In soliton mode locking we assume that the gain, loss and injection terms in the equation are small. Then the target laser waveform has a perturbed soliton component
\begin{equation}\label{eq:psol}
\psi_\text{sn}(t,\tau)=a\sech(a(t-S))e^{ip(t-S)+i\Phi}
\end{equation}
where 
\begin{align}
S(\tau)&=2\int^\tau d\tau'p(\tau')+s\\\Phi(\tau)&=\int^\tau d\tau'(a(\tau')^2-p(\tau')^2)+\varphi
\end{align}
In addition to the strong pulse, the solution of the master equation (\ref{eq:master-inj}) includes a weak continuum component, as well as a response pulse if the target fails to lock to the source. We are interested in the soliton-like pulse component (\ref{eq:psol}) whose properties govern the frequency comb locking.

The slow time dynamics of the pulse parameters, amplitude $a$, phase $\varphi$, frequency $p$, and timing $s$ is obtained by projection of Eq.\ (\ref{eq:master-inj}) on the four NLS adjoint eigenfunctions $\underline{\chi}_k$, $k=a,\varphi,p,s$ \cite{solpert} that gives
\begin{align}
\partial_\tau a&=2g(1-\tfrac\gamma3(a^2+3p^2))a+\bar\sigma(a)-\tilde\psi_\varphi\label{eq:dta}\\
\partial_\tau \varphi&=\tilde\phi_a-\tfrac{p}{a}\tilde\psi_p\\
\partial_\tau p&=-\tfrac43\gamma gpa^2+\tfrac{1}{a}\tilde\phi_s+\tfrac{p}{a}\tilde\psi_\varphi\\
\partial_\tau s&=-\tfrac{1}{a}\tilde\psi_p\label{eq:dts}
\end{align}
where 
\def\vhb#1{\vbox{\hbox{#1}}}
\begin{align}
\tilde\psi_k&=\im\int_{-\infty}^\infty\!\! dt\underline{\chi}_k^*(t)\psi_\text{so}(t)\ ,\\\bar\sigma(a)&=-\im\int_{-\infty}^\infty\!\! dt\underline{\chi}_\varphi^*(t)\sigma(|\psi_\text{sn}(t)|^2)\psi_\text{sn}(t)
\end{align}
\begin{figure*}[htb]
\centerline{\includegraphics[width=8cm]{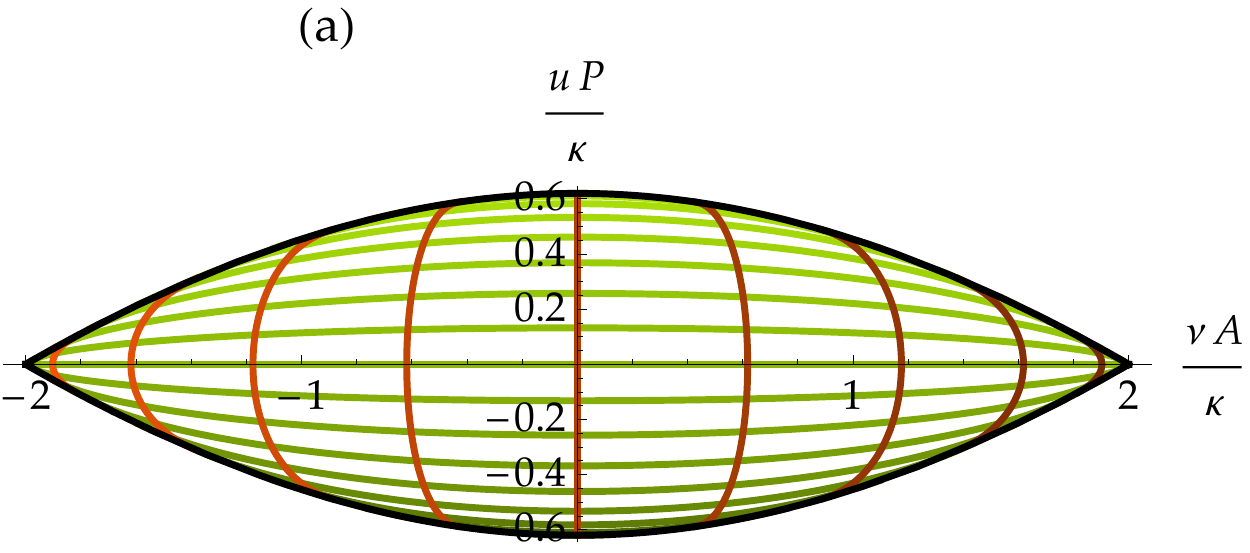}\quad\includegraphics[width=7.2cm]{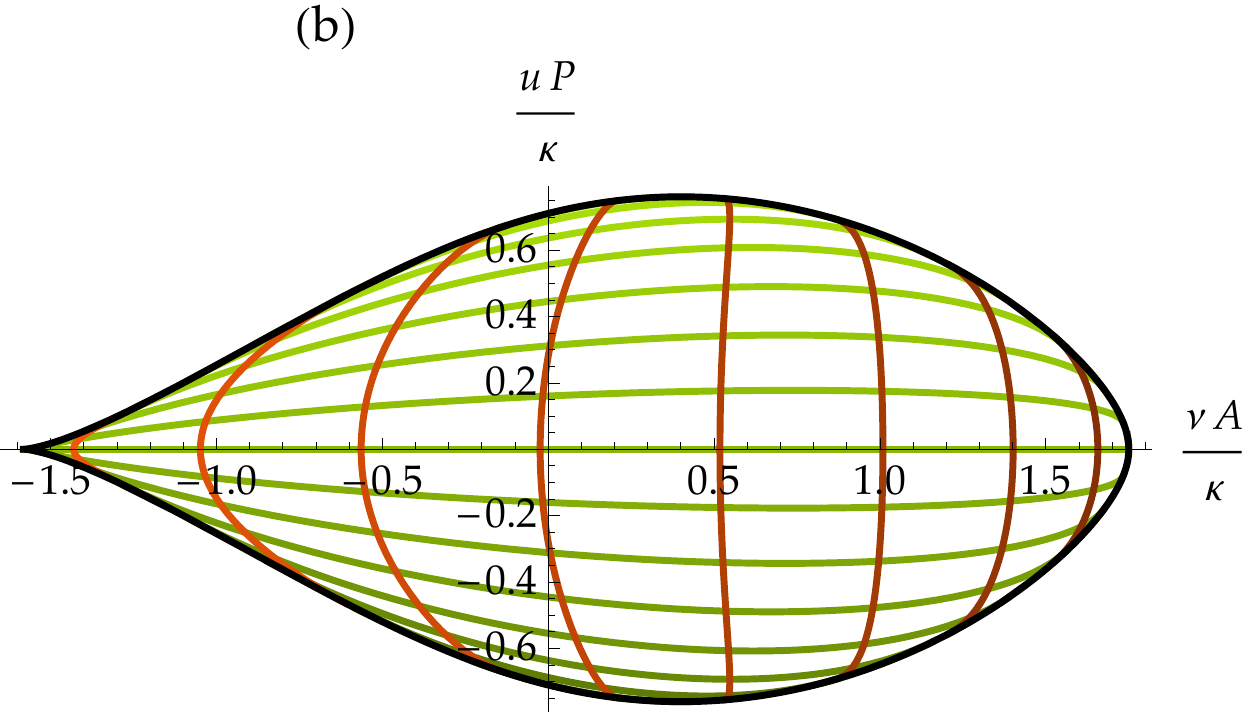}}\centerline{\includegraphics[width=8cm]{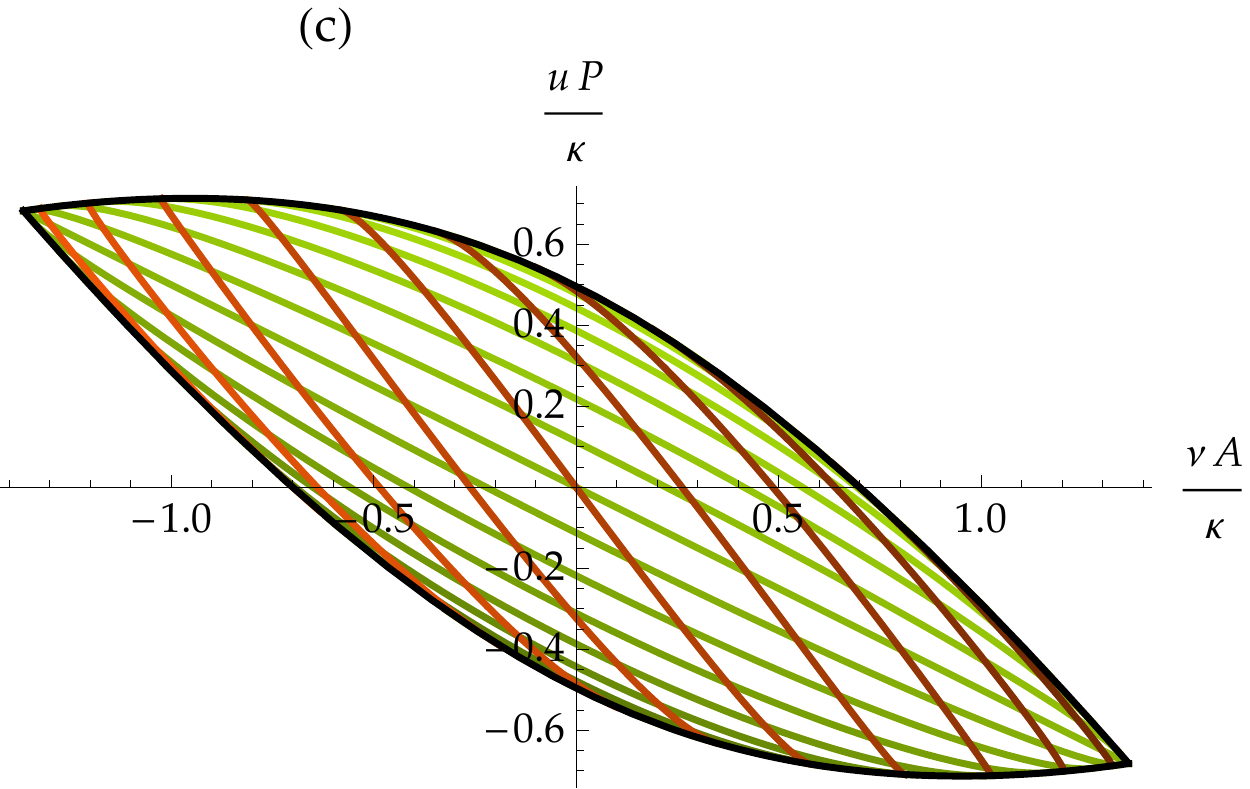}\quad\includegraphics[width=8cm]{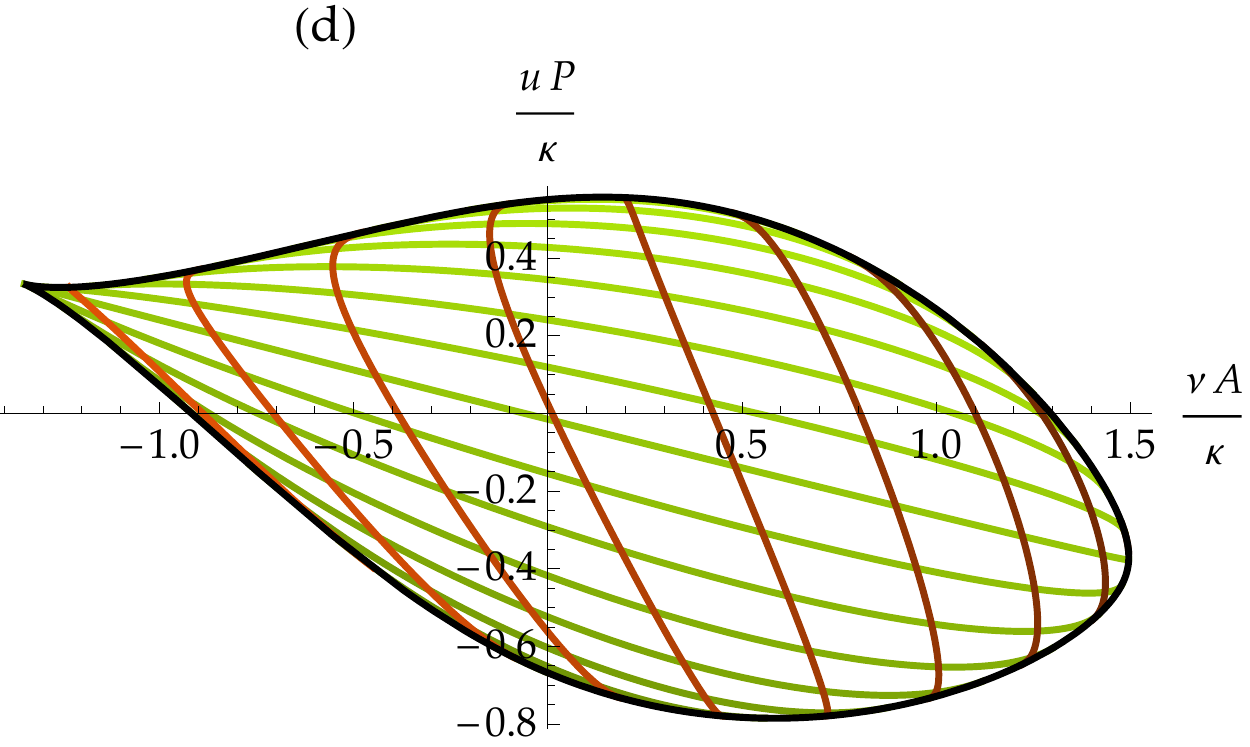}}\vspace{0mm}
\centerline{\quad\includegraphics[width=5.cm]{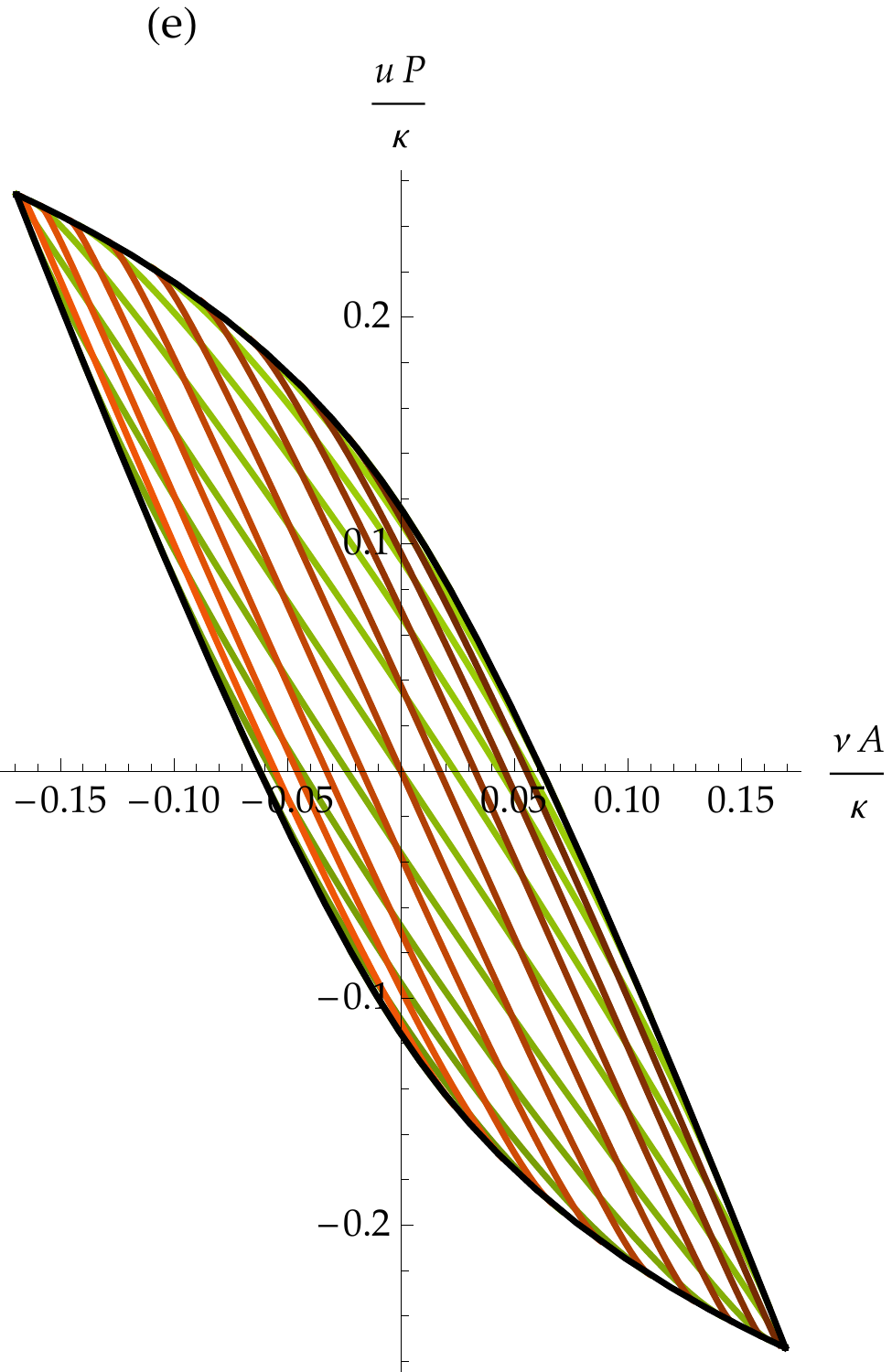}\qquad\includegraphics[width=7cm]{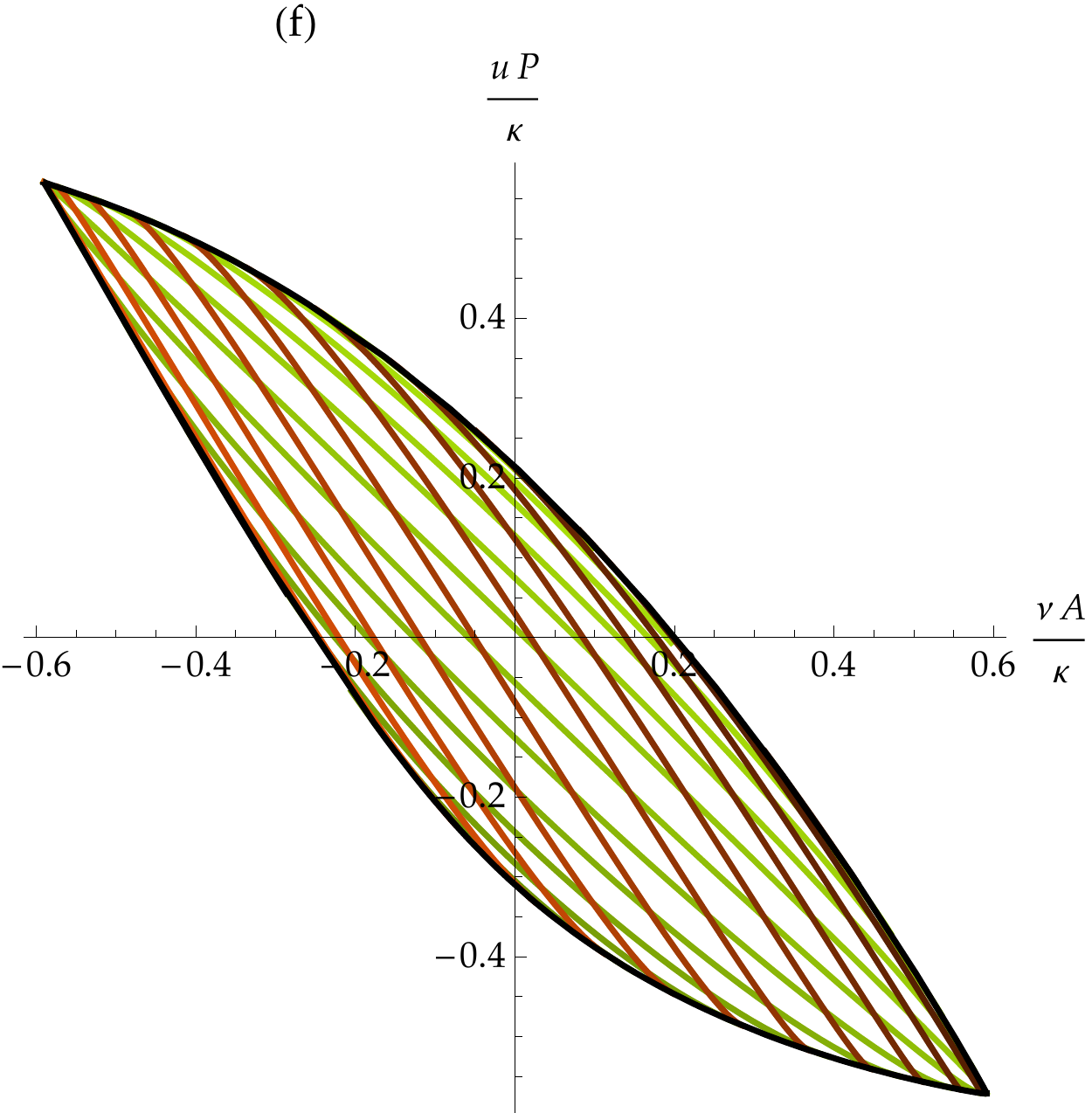}\qquad\quad\vbox{\vhb{\includegraphics[width=12.mm]{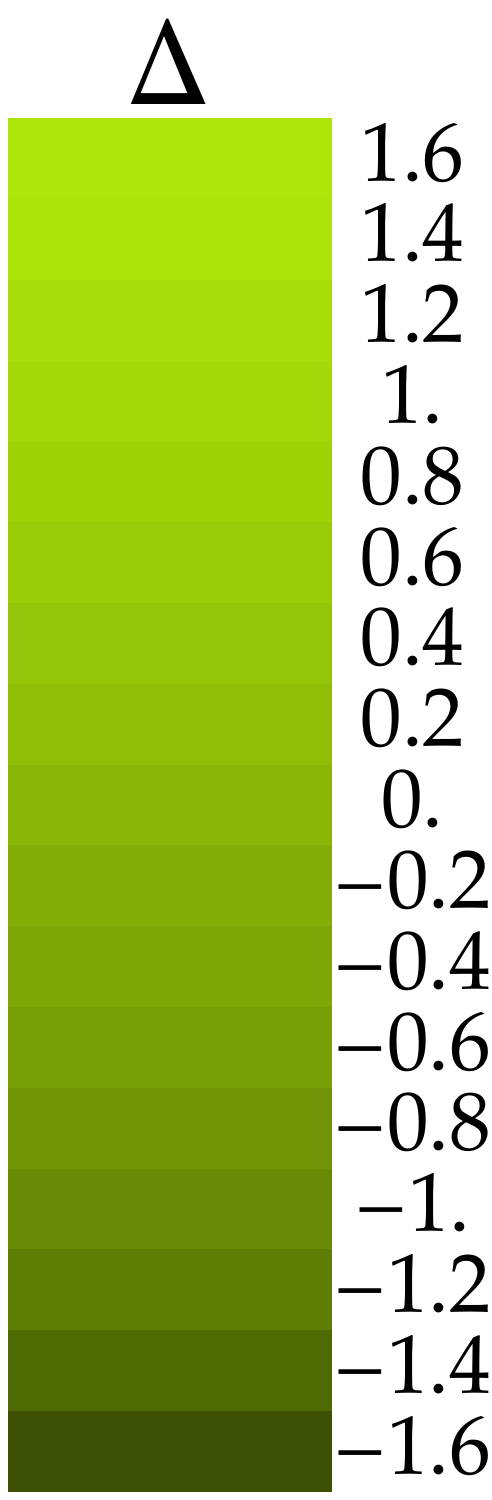}}\vhb{\includegraphics[width=12.mm]{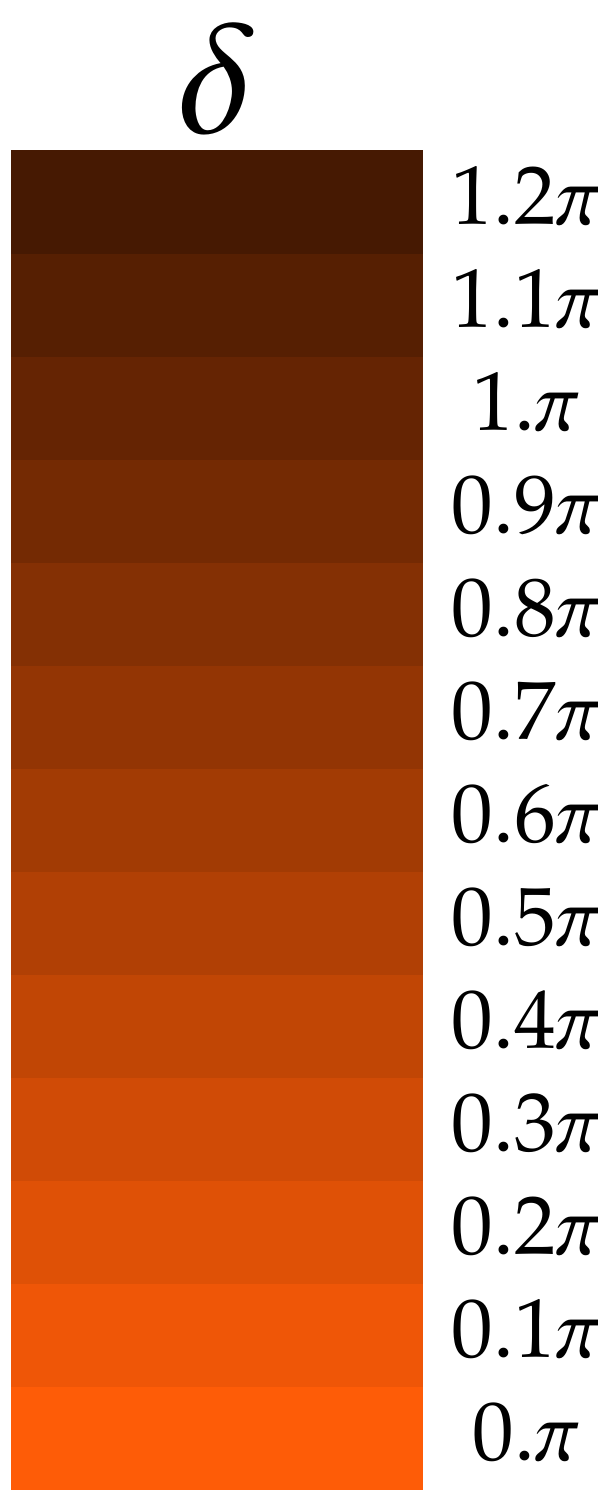}}}}
\caption{\label{fig:ld}Frequency comb locking diagrams for an attenuated soliton source pulse envelope. The horizontal coordinate designates the mismatch $\nu$ between the source and free target frequency comb offsets times the pulse amplitude restoring coefficient $A$ divided by the forcing strength $\kappa$ . The vertical coordinate designates the mismatch $u$ times the pulse frequency restoring coefficient $P$ divided by $\kappa$. All quantities are dimensionless, expressed in natural units as explained in the text.
Points inside the locking domain, whose boundary is the black curve, designate values of mismatches that allow for stable injection locking, where the target comb is pulled and locked to the source comb. Curves of constant source-target time shift in the locked state are shown in green, and constant source-target phase shift curves are shown in orange. The level set values are indicated by hue, coded according to the color bars in the bottom-right part of the figure. The values of source-target frequency difference $\omega$ and chirp $\beta$ are (a) $\omega=0.$, $\beta=0.$, (b) $\omega=0.$, $\beta=0.5$,  (c) $\omega=1.$, $\beta=0.$, (d) $\omega=1.$, $\beta=0.5$,(e) $\omega=3.$, $\beta=0.$, (f) $\omega=2.$, $\beta=.2$
}
\end{figure*}
The seed and target are synchronized if the seed-target phase mismatch $\delta=(\nu_0+\nu)\tau-\Phi$ and timing mismatch $\Delta=u\tau-S$ are locked. It follows that a locked state is a solution of Eqs. (\ref{eq:dta}--\ref{eq:dts}) with $\partial_\tau a=\partial_\tau p=0$, $\partial_\tau\varphi=a^2-p^2$, and $\partial_\tau s=2p$. The seed projections depend on the mismatches so that the injection problem reduces in this regime to the solution of four coupled nonlinear equations for the unknowns $a,p,\delta,\Delta$ given the mismatch parameters $u$ and $\nu$.

The free-running version of equations (\ref{eq:dta}--\ref{eq:dts}) describes mode locked soliton-like pulses with $p=0$, $a=a_s$, where $a_s$ is the solution of
\begin{equation}
G(a_s)\equiv2g(a_s)(1-\tfrac\gamma3a_s^2)a_s+\bar\sigma(a_s)=0
\end{equation}
with free overall phase accumulation rate $\nu_0=a_s^2$.
The question of synchronization is most interesting when the injected signal is weak and the mismatches are small. Here weak injection means that the injected signal is weaker than the intrinsic target laser gain and loss effects (themselves weaker than the the dispersive effects) so that the pulse width and frequency are only slightly shifted from their free running values, that is $\mathring{a}\equiv a-a_s\ll a_s$ and $p\ll a_s$. In this case it is natural use units such that $a_s=1$, so that the locking equations become non-dimensional and linear in $\mathring{a}$ and $p$
\begin{align}
\partial_\tau\mathring{a}&=-A\mathring{a}-\tilde\psi_\varphi\label{eq:dtringa}\\
\partial_\tau \delta&=\nu-2\mathring{a}-\tilde\psi_a\label{eq:dtdelta}\\
\partial_\tau p&=-Pp+\tilde\psi_s\label{eq:dtp}\\
\partial_\tau \Delta&=u-2p+\tilde\psi_p\label{eq:dtDelta}
\end{align}
where $A=-\partial_aG(a_s)$, necessarily positive for pulse stability, and $P=\tfrac43\gamma g(a_s)$. The smallness assumptions used to derive Eqs.\ (\ref{eq:dtringa}--\ref{eq:dtDelta}) imply that $|\tilde\psi_k|\ll A,P\ll1$.

\section{The injection-locked steady state}
Our goal is to map the set of stable injection-locked steady states. In the steady state equations\ (\ref{eq:dtringa}--\ref{eq:dtDelta}) can be further simplified since the direct forcing terms in the phase and timing equations of motion are smaller than the indirect forcing via the amplitude and frequency, as a consequence of a general property of soliton perturbation theory. Specifically, it follows from equations \ (\ref{eq:dtringa},\ref{eq:dtp}) that $|\mathring{a}|\sim|\tilde\psi_\varphi|/A\gg|\tilde\psi_a|$, and therefore the forcing term in equation (\ref{eq:dtdelta}) is negligible; the term $\tilde\psi_p$ is negligible in  equation (\ref{eq:dtDelta}) for analogous reasons.

By the preceding arguments, given the injection waveform and the timing and phase mismatches, a locked steady state is a solution $(\Delta,\delta)$ of the coupled nonlinear equations
\begin{align}
{\nu}&=-\frac{2\kappa}{A}f_\nu(\Delta,\delta)\\
u&=\frac{2\kappa}{P}f_u(\Delta,\delta)
\end{align}
where 
\begin{align}
\kappa f_\nu&=\re\Bigl( e^{i\delta}\int\sech(t)\psi_\text{so}(t-\Delta)dt\Bigr)\label{eq:fnu}\\
\kappa f_u&=\im\Bigl(e^{i\delta}\int\sech(t)\tanh(t)\psi_\text{so}(t-\Delta)dt\Bigr)\label{eq:fu}
\end{align}
displaying explicitly the forcing strength coefficient $\kappa$ of the source waveform.
%It is evident  that if $f_\nu$ and $f_u$ are of $O(1)$ then $\nu$ and $u$ have to be of order $\frac{\kappa}{A}$ for a solution of these equation to exist.
The solution is stable if the eigenvalues of the matrix (omitting zero elements for clarity)
\begin{equation}S=
\begin{pmatrix}
-A&&-\kappa\partial_\Delta f_\nu&-\kappa\partial_\delta f_\nu\\
&-P&~~\kappa\partial_\Delta f_u&~~\kappa\partial_\delta f_u\\
-2&&&\\&-2&&
\end{pmatrix}
\end{equation}
have negative real parts.

The results derived so far hold for weak injection with an arbitrary pulse shape. The precise locking properties depend on the details of the pulse shape through the forcing functions $f_\nu,f_u$; nevertheless, since the forcing functions are defined by overlap intervals, their values are of $O(1)$ for a source pulse shape that matches the free target pulse shape in width and frequency, and is not excessively chirped. We therefore choose the form (\ref{eq:as}), allowing for mismatch $\omega$ in center frequency and $\beta$ in chirp between source and target. Unlike the comb mismatch parameters $\nu$ and $u$, $\omega$ and $\beta$ are pulse parameters whose values are not fixed by the locking dynamics. We further comment below on the different roles played by these two pairs of parameters.

The solution of the injection locking problem is presented in figure \ref{fig:ld} as a set of locking diagrams for different choices of $\omega$ and $\beta$, that display the regions in parameter space where synchronization is stable. In all cases there is a stable locked state when ${\nu}=u=0$, while for large ${\nu}$ or $u$ there are no stable locked states. Frequency comb locking is made easier by stronger injection, and hindered by stronger restoring coefficients, so that the mismatches are multiplied by the restoring coefficients ($\nu$ by $A$ and $u$ by $P$) and divided by $\kappa$ in the locking diagrams; the physical implications of the scaling are analyzed in the conclusions section below. The locked and non-locked regions in the ${\nu}$-$u$ plane are separated by a curve where at least one the eigenvalues of $S$ is zero, shown in black in the locking diagrams. It follows that at the locking boundary point the Jacobian matrix $\Big(\begin{smallmatrix}\partial_\Delta f_\nu&\partial_\delta f_\nu\\\partial_\Delta f_w&\partial_\delta f_w\end{smallmatrix}\Big)$ is singular, so that at generic boundary points there is a saddle-node bifurcation, where  stable locked solutions connect with unstable locked solutions on reaching the locking domain boundary from the interior. It also follows that the constant $\Delta$ and constant $\delta$ curves, shown in Figs.\ \ref{fig:ld} in green and orange hues (respectively) meet the boundary tangentially.

In the non-chirped locking diagrams the $\Delta=0,\,\delta=0$ and $\Delta=0,\,\delta=\pi$ are mapped to corners on the locking domain boundary, since the Jacobian vanishes there identically. The corners are not structurally stable, and when the source is chirped one of the corners becomes smooth and the other a cusp. The boundary of a typical locking diagram is therefore smooth except for one cusp.

\begin{figure}\includegraphics[width=7cm]{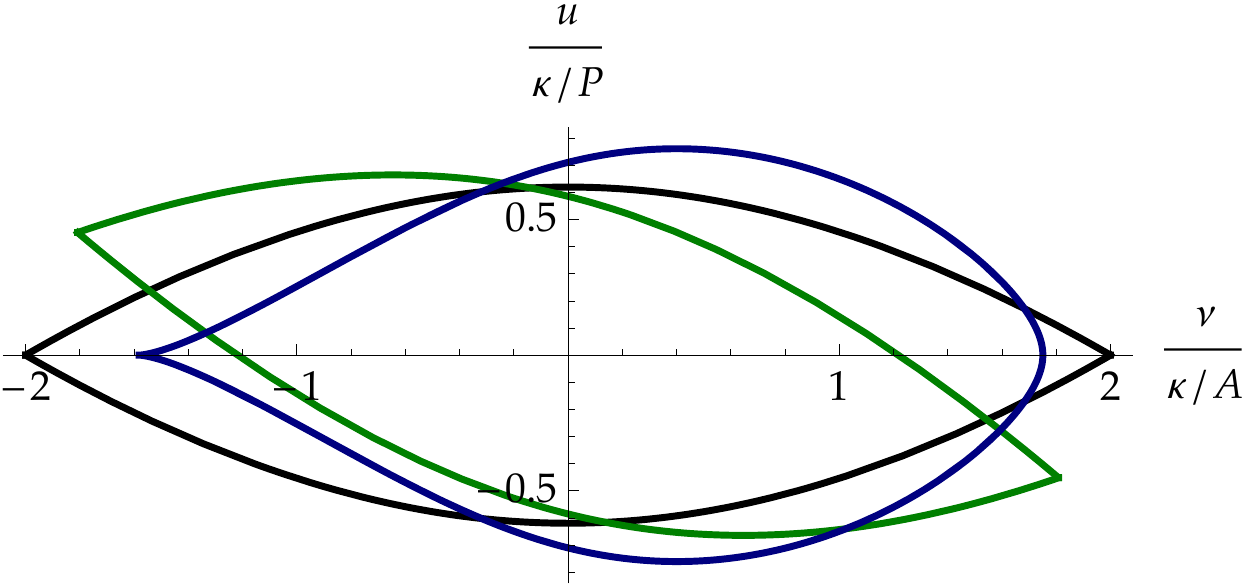}
\caption{\label{fig:nm}Boundary of locking region for $\omega=0.,\,\beta=0.$ (black), $\omega=0.5,\,\beta=0.$ (green), and $\omega=0.,\,\beta=0.5$ (blue). It is evident that moderate source chirp and source-target frequency shift improves locking properties for some comb mismatch values. The axes are as in Fig.\ \ref{fig:ld}
}
\end{figure}

The most evident conclusions from the locking diagrams is that the locking region shrinks with both increasing frequency mismatch and increasing chirp. This results is to be expected since the overlap of the source and target pulses decreases when there is relative frequency or chirp, and the injection efficiency decreases. When $\omega$ and $\beta$ become large the area of the locking domain decreases fast to zero---as a power law for large $\beta$, and exponentially for large $\omega$.
Nevertheless, the shrinking is not monotonic, and some areas of the ${\nu}$-$u$ plain are more susceptible to locking with moderate frequency mismatch and/or moderate chirp, see figure \ref{fig:nm}. This is made possible by adjustment of the locked phase shifts $\Delta$ and $\delta$ to compensate for greater mismatches. For non-zero $\omega$ the source waveform has a time-reversal odd component, so that the values of $f_\nu$ and $f_u$ become closer, and the admissible ${\nu}$ and $u$ intervals become correlated; the correlation becomes more pronounced for larger frequency mismatch. 

The locking diagrams display some discrete symmetries. First, for any locked state ${\nu},w$ with shifts $\Delta,\delta$, there is an opposite solution $-{\nu},-w$
with shifts $\Delta,\pi+\delta$; at most one of these solutions is stable. For a time-reversal even source envelope the following identities hold
\begin{align}
&f_\nu(\Delta,\delta;\omega,\beta)=f_\nu(\Delta,-\delta;-\omega,-\beta)=\nonumber\\
&\qquad f_\nu(-\Delta,\delta;-\omega,\beta)=f_\nu(-\Delta,-\delta;\omega,-\beta)\\
&f_w(\Delta,\delta;\omega,\beta)=-f_w(\Delta,-\delta;-\omega,-\beta)=\nonumber\\
&\qquad-f_w(-\Delta,\delta;-\omega,\beta)=f_w(-\Delta,-\delta;\omega,-\beta)
\end{align}
It follows that the locking diagrams for all $\omega,\beta$ values are determined by those with both non-negative: The locking diagram for $-\omega,\beta$ is obtained from that of $\omega,\beta$, by horizontal reflection and replacement of $\Delta,\delta$ by $-\Delta,\delta$, and the diagram for $\omega,-\beta$ is obtained from that of $\omega,\beta$ by a $\pi$ rotation around the origin and replacement of $\Delta,\delta$ by $-\Delta,\pi-\delta$. In particular, the locking boundaries of the $\beta=0$ diagrams are symmetric with respect to a half turn around the origin, and those of the $\omega=0$ diagrams are symmetric with respect to horizontal reflection.

\section{Conclusions}
The main result of this work is a precise mapping of the locking of a passive mode locked laser frequency comb to that of an external pulse train. We focused on the weak injection regime where this effect is most striking. In contrast with cw synchronization, frequency combs are specified by two frequencies, that have to be locked together, making the locking problem two-dimensional.
The difference between the two combs was parametrized by the mismatch $\nu$ between the comb offsets and the mismatch $u$ between the comb spacings. The mismatches should be comparable or smaller to the forcing strength $\kappa$ divided by the restoring coefficients $A$ and $P$ (respectively) of the target laser dynamics.

The locking diagrams show the synchronization behavior using dimensionless quantities: $u$ is expressed as multiples of $a_s$, the inverse of the free target pulse width, and $\nu$, $A$, and $P$ as multiples of the dispersion frequency---the dispersion coefficient $\beta$ times $a_s^2$ times the group velocity---that is typically much slower than $a_s$. The injection-restoration ratios $\frac{\kappa}{A}$, $\frac{\kappa}{P}$ are a measure of the strength of the injection, here assumed small. For weak injection therefore, the repetition rate of the source and the target should be matched up to a fraction of the pulse bandwidth, and the offset frequencies have to be matched up to a fraction of the slow frequency unit. If $\kappa$ is increased, the locking domain becomes larger proportionately, until the forcing is no longer weak, when nonlinear  effects not studied here may prevent further growth of the locking domain. 

Practically, the laser repetition rates can be tuned by changing the cavity length, and the pulse bandwidth makes the admissible $u$ interval large enough for passive control. On the other hand, the natural frequency scale of the offset mismatch $\nu$ is much smaller, and the $\nu$ tolerance is accordingly more limited. To understand the locking of pulse phase it should be recalled that $\nu$ is actually defined modulo $\frac{2\pi}{\tau_R}$, where $\tau_R$ is the repetition rate, so that a physical $\nu$ corresponds to a horizontal lattice in the locking diagrams. If $\nu$ takes a random value between $-\frac{\pi}{\tau_R}$ and $\frac{\pi}{\tau_R}$ (say), there is a high probability of locking if the lattice spacing $\frac{2\pi}{\tau_R}$ is of the order of or smaller than the range of offset mismatched in the diagram, and low probability otherwise. In fiber lasers with $0.5\,\text{ps}$ pulses and standard fiber dispersion the slow frequency scale is about $1\,\text{MHz}$---much smaller than normal repetition rates of about $50\,\text{MHz}$. It is therefore difficult to obtain small enough $\nu$ by passive control, and for this reason a sophisticated active stabilization of $\nu$ was implemented in \cite{kg12}  to achieve synchronization. 

The amplitude and frequency restoring coefficients $A$ and $P$ are hard to measure directly. $A$ can be estimated by the relaxation rate of the gain medium, which can vary by many orders of magnitude between $10^2\,\text{sec}^{-1}$ in Erbium and $10^9\,\text{sec}^{-1}$ in diode lasers. $P$ is equal to the product of the overall differential gain $g$ and the gain bandwidth attenuation factor $\gamma a_s^2$. $g$ can be estimated from cw operation where it typically varies between $10^6\,\text{sec}^{-1}$ and $10^8\,\text{sec}^{-1}$, while $\gamma$ is inversely proportional to the square of the bandwidth so that $\gamma a_s^2$ varies from about $10^{-2}$ for highly chirped pulses to 1 for bandwidth limited pulses. Weaker restoring coefficients allow for larger pulling of $a$ and $p$ from their free-running values, and therefore the $\nu$ locking interval is larger for slow gain media and the $u$ locking interval is larger for slower differential gain and larger gain bandwidth.

An important conclusion from this discussion is that the locking properties depend strongly on the pulse duration and the group velocity dispersion. For fixed injection efficiency $\kappa$ and restoring coefficients, the admissible $\nu$ interval is proportional to $\beta^2a_s^4$ and the admissible $u$ interval is proportional to $\beta a_s/\gamma$. It follows that injection locking is particularly well-suited for high dispersion ultrafast lasers.

Unlike the mismatch of comb parameters that acts as a barrier to synchronization, the matching of source and target pulse shapes affects the locking properties through the overlap integrals in equations (\ref{eq:fnu}--\ref{eq:fu}). It follows that the most relevant are the matching of the central frequency, chirp, and width of the pulses. Of these, the locking is most sensitive to frequency mismatch, since the overlap integrals drop sharply when the bands of the source and target do not match. A subtler effect is the deformation of the locking domain to an elongated shape, so that there is an `easy' direction in the $\nu$-$u$ plane where relatively large mismatches allow injection, which is not possible with comparable mismatches in other directions. In practice the frequencies are naturally matched if the two lasers use the same gain mechanism.

The effect of relative chirp is to decrease the effective source bandwidth available for locking, so the area of the locking domain shrinks more slowly for large chirp than as for large frequency mismatch.
The example of chirp also highlights the difference between the role of comb parameter and pulse parameter matching. Whereas for a given chirp there is always an area of the $\nu$-$u$ plane that allows locking, no choice of pulse shapes lead to synchronization if the offset frequency mismatch $\nu$ is significantly larger than $\kappa/A$ in natural units. These conclusions are in agreement with the experimental achievement of injection locking by tuning the chirp of the source pulse in \cite{kg12}.

We did not study in detail the effect of pulse-width mismatch between source target, but  the effect is likely to be similar to that of chirp, as a gradual shrinking of the area of the locking domain, but without the breaking of the discrete symmetries. The study of the effect of noise which is liable to change some qualitative properties  \cite{pik,moh2,crit-th,crit-exp} is beyond the scope of this work. The results derived here have natural implication in threshold phenomena in rational pulse injection locking and the resulting repetition rate multiplication that need to be studied separately.

\emph{Acknowledgments} J Reiner and E Olami participated in early stages of this project. The project was supported by the Israel Science Foundation grant {1002/07}, US-Israel Binational Science Foundation grant 2010442,  D.K. was supported by an Australian Research Council Future Fellowship (FT110100513)..

\end{document}